# Towards SocialVR: Evaluating a Novel Technology for Watching Videos Together

Mario Montagud*[1,2], Jie Li[3], Gianluca Cernigliario[1], Abdallah El Ali[3], Sergi Fernández[1], Pablo Cesar[3]

*Corresponding author

[1] i2CAT Foundation, Barcelona (Spain)

[2] Universitat de València (UV), Spain

[3] Centrum Wiskunde & Informatica, Amsterdam (The Netherlands)

{mario.montagud, gianluca.cernigliario, sergi.fernandez}@i2cat.net; {Jie.Li, Abdallah.El.Ali, P.S.Cesar@cwi.nl}@cwi.nl

*Abstract* — Social VR enables people to interact over distance with others in real-time. It allows remote people, typically represented as avatars, to communicate and perform activities together in a join shared virtual environment, extending the capabilities of traditional social platforms like Facebook and Netflix. This paper explores the benefits and drawbacks provided by a lightweight and low-cost Social VR platform (*SocialVR*), in which users are captured by several cameras and reconstructed in real-time. In particular, the paper contributes with (1) the design and evaluation of an experimental protocol for Social VR experiences; (2) the report of a production workflow for this new type of media experiences; and (3) the results of experiments with both end-users (N=15 pairs) and professionals (N=25) to evaluate the potential of the *SocialVR* platform. Results from the questionnaires and semi-structured interviews show that end-users rated positively towards the experiences provided by the *SocialVR* platform, which enabled them to sense emotions and communicate effortlessly. End-users perceived the photo-realistic experience of *SocialVR* similar to face-to-face scenarios and appreciated this new creative medium. From a commercial perspective, professionals confirmed the potential of this communication medium and encourage further research for the adoption of the platform in the commercial landscape.

# 1. Introduction.

Virtual Reality (VR) is the experience in a simulated interactive virtual space. It provides synthetic sensory feedback to users' actions that can physically and mentally immerse the users [49]. Social VR is one type of VR systems that allows multiple users to join a collaborative virtual environment and communicate with each other, usually by means of visual and audio cues [10, 11, 34]. The virtual space can be a computer-generated 3D scene or a 360º scene captured by an omnidirectional camera. Each user is represented as a computer-generated avatar [17, 22, 30, 45, 51] or, in recently proposed systems, using a virtual representation based on live capture [21, 25]. Depending on the system, the user's virtual representation can also interact with the virtual space, for example by manipulating virtual objects, controlling the appearance of the space, or controlling the playout of additional media.

The interest in Social VR has grown in the last few years, and particularly in the last few months. Many platforms have been developed[1], enabling richer interaction between remote users than traditional social platforms such as Facebook, YouTube and Netflix. This paper introduces a lightweight Social VR platform (*SocialVR*), which provides hyper-realistic experiences based on real-time capturing and reconstruction of the users. Based on a relevant Social VR scenario [20], which is remote users watching content together, the paper explores the new production process needed for Social VR content and presents an evaluation of the overall experience from both the end-user and commercial perspectives. It thus reports a holistic evaluation of the full workflow from creation to experience delivery.

There are two key aspects that make our *SocialVR* platform outstanding from the state-of-the-art. First, our *SocialVR* platform enables photo-realistic user representations, unlike most of the existing solutions in which users are represented as puppet-like avatars (e.g., Sansar, AltspaceVR, and Facebook Horizon). Second, our platform is lightweight, low-cost and uses off-the-shelf hardware, unlike other platforms that require high-end hardware and a fast Internet connection for achieving high quality real-time 3D reconstructions and providing realistic representation of users [4, 16, 42]. In particular, our platform enables real-time communication capabilities based on 3D volumetric representations, which are reconstructed from a number (e.g. typically 4) of affordable RGB-D cameras (e.g. Kinect v2 sensors) surrounding the participants.

The research contributions of this paper in the Social VR field are detailed next. As a first contribution, the paper presents a protocol for qualitatively evaluating Social VR experiences, based on shared media consumption. The experimental protocol is identified, and validated by running a test based on comparing shared video watching experiences using three setups: (1) face-to-face scenario; (2) use of Facebook Spaces; and (3) use of a video-based Social VR system. The protocol worked successfully, showing users' preferences towards photo-realistic representations.

---

This supports the decision on building an own fully controllable and enhanced *SocialVR* platform, based on photo-realistic volumetric representations. As a second contribution, the paper reports on the evaluation of the novel Social VR experience with end-users (*N=30*). The findings from the experiment using the *SocialVR* platform prove that end-users rated positively towards the developed *SocialVR* platform and provided experience, which enables them to sense emotions, communicate naturally, and experience social presence [41]. As a third contribution, the paper provides an in-depth analysis of the added value of Social VR, in terms of commercial opportunities. The findings from the Social VR experiment with VR professionals (*N=25*) confirm the potential of this medium, revealing use cases and business opportunities, and encourage further work towards the adoption of the platform in the commercial landscape. As an added on, the paper provides insights about the technology and production workflow that are needed in order to create novel hyper-realistic Social VR experiences. Figure 1 shows an example of users interacting with our *SocialVR* platform.

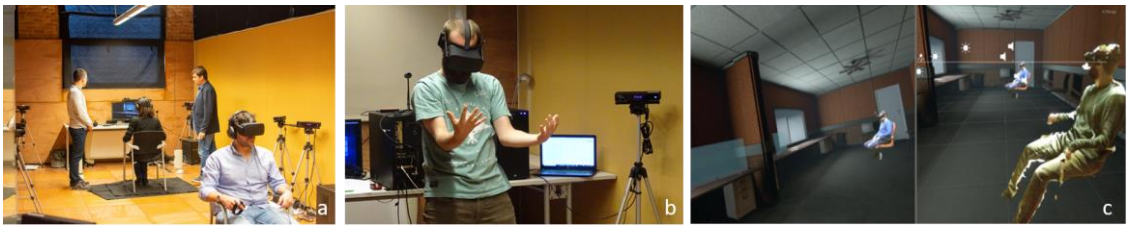

*Figure 1. Users Experiencing our SocialVR platform: (a) The lab and system setup; (b) A user recognizing his hands (self-representation in VR); (c) Two users integrated in a Social VR scenario*

## 2. Related Work

The interest for Social VR systems dates back to the late 90s [10, 11, 17, 38, 57]. Recently, Social VR has been increasingly attracting attention both in commercial applications and academic research. Current VR platforms, such as Sansar, AltspaceVR, Mimesys and Facebook Horizon, seek to include Social VR features in their systems [22, 30, 45, 50, 51, 56]. This section overviews three different aspects related to Social VR: shared media consumption, systems, and evaluation.

### 2.1. *Shared Media Consumption in Social VR*

Traditionally, family members and friends gather at a common location for watching media content together (e.g. TV, videos, photos…). This social gathering enables people to interact and share emotions, thus contributing to an increased engagement and social bonds between them. However, nowadays, it is not always possible for people to meet at the same place. Due to this, many technologies have been developed to enable people to remotely share media consumption (e.g. by means of text, audio and/or video chat channels). In the particular case of TV content, the Social TV concept has been generally adopted [5, 6].

Research on Social TV has attracted attention in the last decade. Some example works focused on: analyzing the advances in Social TV and categorizing the existing developments [6]; studying the appropriateness of different chat modalities [24]; determining the impact of delays [19]; and assessing the interest in these scenarios [3]. Likewise, many lab-controlled [19, 39] and in-home [40] studies have shown the benefits provided by Social TV mainly in terms of togetherness, intimacy and improved relationships.

Given the benefits provided by both co-located and remote shared media consumption, the research community recently started to explore how to support them through Social VR platforms. Facebook Horizon[2] and Sansar[3] are two examples of Social VR platforms that enable shared media consumption, by representing users as avatars. Rothe et al. [46] proposed guiding and interaction techniques to efficiently support remote shared media consumption of 360º videos on Head Mounted Displays (HMDs). McGill et al. [40] showed that the adoption of HMDs in conjunction with RGB-D cameras for users' capturing can lead to an increased engagement, feeling of immersion and enjoyable embodied telepresence compared to video conferencing tools. Gunkel et al. [21] developed a video-based Social VR platform, also mainly focused on shared media consumption. In that platform, the users are captured photo-realistically by a single RGB-D camera (Kinect), and the shared VR scenario is represented as a 360º static image.

## 2.2. Building Social VR systems based on volumetric media

Video technologies are constantly evolving towards higher quality and more efficient and immersive formats. In this context, the research community has been focusing on the development of novel systems to represent, compress and transmit volumetric (3D) videos, both for artificially generated content (e.g. Computer Graphics (CG)) and natural content (e.g. realistic furniture and humans). Natural content needs to be realistic in order to accurately represent humans in Social VR, thus providing enjoyable experiences. Two representation formats are commonly used to render natural content: meshes [37] and point clouds [12, 33].

Together with this evolution, new technologies for real-time 3D video capturing, transmission and presentation are being developed. This has brought the concept of *holoportation*, allowing a real-time 3D projection of remotely captured videos, like in the Microsoft system for HoloLens [42]. In such scenarios, given the real-time requirements, the representation and processing of the natural content become crucial. New ways to acquire, compress and transmit volumetric content are thus being explored, for both meshes and point clouds. Meshes are evolving towards the concept of Time Varying Meshes (TVMs) that provide a mesh based volumetric representation of the natural content, captured and reconstructed in real time. The compression of TVMs has been widely explored in the past two decades, considering both spatial and temporal redundancy [36]. Point clouds are currently considered the most appropriate 3D representation of natural content and, consequently, they are attracting attention within Moving Pictures Experts Group (MPEG) [7], where a first standardization process for this volumetric format has been initiated, called Point Cloud Compression (PCC) [48].

---

Video-conferencing is then evolving, being able to support emerging immersive video formats. This allows the creation of Virtual/Augmented Reality (VR/AR) meeting systems, where 3D and 2D video, together with interactive CG content, are represented in 3D spaces [13]. The users of these systems are able to meet, collaborate and work remotely [2], having the feeling of being immersed in the same room. An example of hybrid Social VR platform is DataSpace, developed by IBM [4], which allows the creation of collaborative VR/AR spaces. In such scenarios, users are able to collaborate and interact remotely in a 3D environment and play with different types of content. Likewise, a volumetric display system for holographic conferencing through VR/AR spaces has been developed by Mimesys[4]. It relies on the use of Intel RealSense cameras and Magic Leap One headsets. Other Social VR platforms, apart from the interaction, have also focused on the representation of the users in the virtual environment. An example is the work of Fairchild et al. [16], where a free viewpoint video (FVV) system is provided to capture, deliver and render the users in real-time.

### 2.3. *Evaluating Social VR*

Instead of watching a film or playing a game together through a screen, Social VR can be experienced as if the viewers were actually co-present in the same space. Many works have evaluated the impact of system design factors, such as the avatar appearance and its behavioral realism, on user's quality of experience and social interactions in Social VR. We provide hereafter an overview of the methods and findings of some recent studies.

Li et al. [34] proposed an experimental protocol and a questionnaire for measuring experiences in Social VR. Using photo sharing as a use case, they compared quality of interaction, social meaning and presence/immersion in face-to-face, Skype and Social VR scenarios. The results of the experiment suggest that the construct of the Social VR questionnaire is valid with high internal reliability. The questionnaire can be generalized to other use cases.

Many studies identified the importance of embodiment for providing immersive experiences [28]. Heidicker et al. [22] compared different user representations. The results indicate that motion-controlled avatars, even only having head and hands visible, produced an increased feeling of co-presence and behavioral interdependence. Motion-controlled full body avatars lead to an increased sense of presence. Roth et al. [45] studied whether realistic body movements of full-body avatars could compensate for missing facial expression and eye gaze cues. The results indicate that social interactions tend to be impeded with non-realistic avatars, but the absence of important behavioral cues, such as gaze and facial expression, can be partially compensated by realistic body movements. Smith and Neff [51] compared the audio-visual communication between two users completing a task in three conditions: (1) face-to-face; (2) Social VR with embodied avatars that have an eyebrow ridge and nose, but no other facial features; and (3) Social VR without visible avatars, but only virtual hands. They concluded that embodied avatars provide a high level of social presence with conversation patterns that are very similar to face-to-face interaction.

---

[4] https://www.mimesysvr.com. Last access in April 2021

In addition, many other studies investigated how the realism of the virtual representations influence the social interaction. Latoschik et al. [30] explored the effect of avatar realism on self-embodiment and social interactions in Social VR. Realistic avatars were rated significantly more human-like and evoked a stronger acceptance in terms of virtual body ownership. Similarly, Waltemate et al. [56] found that personalized avatars significantly increase the sense of body ownership, presence, and dominance compared to other two explored avatars.

The evaluation in these related works was conducted mainly using questionnaires, interviews, and some of them analyzed the verbal and non-verbal cues from the video recordings. The protocol used by Smith and Neff [51] is very complete, and served as an inspiration to our protocol design.

## 3. Designing and Testing an Evaluation Protocol for Social VR

As a preparation for the implementation of our *SocialVR* platform, we designed and evaluated a protocol for evaluating a shared video watching experience when using two existing Social VR systems. The goal is to understand which type of Social VR system can better support social experiences, such as watching and talking about a movie trailer, and whether the data collected through questionnaires, interviews, and video annotations are valid.

### 3.1. *The Protocol Test*

We compared video watching experiences in three conditions: (1) Facebook Spaces (FS), with a half-body cartoon avatar with a customizable look; (2) A video-based Social VR system (VB) [21], with 2D photo-realistic representation captured in real-time using a Kinect sensor; and (3) Face-to-face (F2F) scenario as a benchmark. In the first two conditions, the two participants were wearing an Oculus Rift HMD and noise-cancelling headphones, and sitting on a chair fixed to the floor in two separate rooms (see Figure 2). They were teleported to a virtual room represented using the same 360º image of an office, and watched movie trailers together on a virtual screen. They interacted with each other through audio and visual interaction channels. In the F2F scenario, the two participants were physically sitting together in the same room with a TV screen in front of them (see Figure 3).

Sixteen acquaintance pairs of participants (e.g., friends, colleagues) were recruited (17f, M=31.1, SD=7.4) to test the protocol. We selected three trailers of action/science fiction movies with approximately the same duration (about 2,5 minutes) and number of views on YouTube. Some previous research has shown that choosing movie trailers with the length of 2,5 minutes are sufficient in measuring brain responses to the trailers [9] and video Quality of Experience (QoE) [35]. The three conditions were counterbalanced and followed a balanced Latin Square design. The video played in each condition was randomized. So, each pair of participants experienced all three conditions, watching a different movie trailer in each condition.

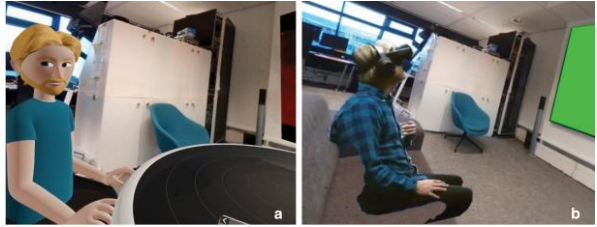

*Figure 2. (a) One user's view in the Facebook Spaces system (FS); (b) One user's view in the video-based Social VR system*

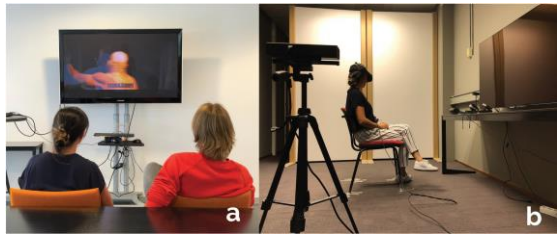

*Figure 3. The protocol test setup for the (a) F2F and (b) Social VR conditions*

Both participants' audio and visual interactions and body movements were recorded using a webcam in all three conditions. After each condition, participants were asked to fill in a social VR questionnaire from Li et al. [34], which was developed based on photo sharing experiences in social VR.We used the same items as in the original questionnaire, but adapted them to the shared video watching experience. The questionnaire covers three main factors of user experience (see Appendix A): the quality of interaction (QoI), the social meaning (SM), and the sense of presence/immersion (PI). After completing all three conditions, a semi-structured interview was conducted with the two participants.

### 3.2. *Results of the Protocol Test*

**Questionnaire**. Figure 4 shows the box plots of the subjective scores collected via the questionnaire for the QoI, SM and PI factors. PI was only assessed in the two Social VR conditions. Via a Friedman rank sum test and Wilcoxon-Mann-Whitney test, we found a significant effect of the system condition on QoI ($\chi^2(2)$=17.7, p<.001), with FS < F2F (p<.001, r=0.57), FS < VB (p<.01, r=0.36) and VB < F2F (p<.01, r=0.37). By performing the same tests, we also found a significant effect on SM ($\chi^2(2)$=10.9, p<.01), with FS < F2F (p<.001, r=0.51) and FS < VB (p<.02, r=0.31).

**Semi-structured interviews**. The audio recordings of the semi-structured interviews were transcribed and coded by two researchers, following an open coding approach [53]. From the coded transcripts, several themes emerged, which we discuss below. The sixteen pairs of participants are labeled P1A(B) - P16A(B).

Almost half of the participants (47%) expressed concerns that the avatars in the FS system were not realistic. Some of the participants (22%) explained that the facial expressions were limited, which influence the communication. Others (19%) mentioned that the body language was also missing when having avatars. Therefore, they felt that the avatar was not helpful in supporting communication (P4B: *"We didn't look at each other while watching the trailer".*). The user representation in the VB system was believed by some participants (28%) to be more personal and natural, compared with the FS system. With the photo-realistic representation, the participants were able to interpret the emotions of each other (P9A: *"If you looked into each other in the VB system, you can somewhat interpret the emotions".*). Some participants (41%) felt that the eye contact was not necessary, while others (38%) were bothered by the blocked eye contacts by the HMD. Another difference between the two systems reported by the participants was the controller. They felt the controllers were difficult to use (16%) in the FS system, and did not want to hold the controllers all the time (22%). Half of the participants (50%) preferred the VB system for activities such as watching a movie, and others (34%) recommended the FS system for gaming use cases.

Around 25% of the participants said the quality of VR environments was acceptable. For the virtual environment of the VB system, some participants (34%) felt anxious "sitting there" (P9A: *"I felt I must sit still. If I moved, I could fall down".*). 38% of the participants mentioned that they did feel presence in both VR environments. However, the presence feeling was influenced by the use of static 360° scenes, which were perceived as blurry in some cases. 22% of the participants suggested providing better body representations and enabling automatic gesture recognition.

**Participants' interaction and behavior**. Two researchers manually annotated the time two participants spent talking and looking at each other from the recorded videos. Figure 5 shows the box plots for the percentage of time spent looking at each other and talking to each other. The data were analyzed using the repeated-measures ANOVA and multiple comparison test. We found a significant effect of condition on the percentage of time spent talking at each other, with F2F < VB (p < .003).

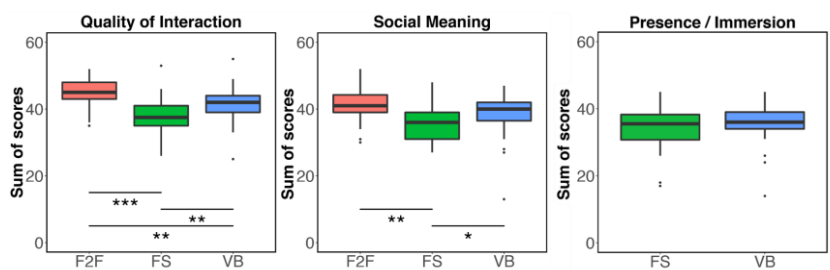

*Figure 4. Box plots of the questionnaire subjective scores for F2F, FS and VB conditions*

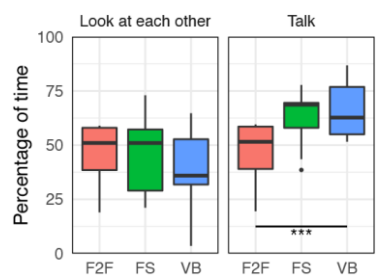

*Figure 5. Box plots of the percentage of time spent talking to and looking at each other*

### 3.3. *Reflection on the Protocol Test*

Based on the test of the protocol, we found that the collected data through the questionnaire, interviews and video recordings exhibited sufficient results. Based on the results, we are able to compare the Social VR experiences in two different Social VR systems (i.e., FS and VB) and conclude that the VB system performed better than the FS system in terms of both QoI and SM, and that VB tends to encourage participants to spend more time talking with each other.

This suggests that the designed evaluation protocol for Social VR is appropriate, and thus it is going to be used for the evaluation of the newly developed Social VR experience, combining a new *SocialVR* platform and created VR story. Apart from the protocol, the results prove that both Social VR systems used in the protocol test supported real-time communication between participants, and suggest that the system with photo-realistic user representation provides a more engaging experience than FS with cartoon avatars. This is inline with existing findings in literature, concerning the importance of having realistic avatars [17, 22, 30, 45, 51].

## 4. Building a Hyperrealistic Social VR Experience

This section describes the lightweight and hyperrealistic *SocialVR* platform that has been used for evaluating the Social VR experience. In addition, it provides insightful details about the creation of a professional VR story for this new kind of experiences.

### 4.1. *SocialVR Platform and Setup*

Based on the insights from the protocol test, it was decided to use a new Social VR platform (*SocialVR*) where we could have full control over the technology and experience. Thanks to this platform, users are able to see each other's gestures and communicate through audio in real-time in a shared VR environment. Interestingly, it supports a real-time capturing, encoding, distribution and rendering of volumetric videos representing the end-users, even with self-views. This represents an outstanding feature compared to state-of-the art solutions (Section 2).

This section presents an overview of the *SocialVR* platform. Its architecture and the streams exchanged between its components are shown in Figure 6. The main parts of the *SocialVR* platform, including technical and implementations details, are described in the following subsections.

**4.1.1. Capturing & Reconstruction**. To enable photo-realistic and fluid volumetric representations of users in the *SocialVR* experience, a real-time video capturing and reconstruction sub-system has been integrated, based on the work by Alexiadis et al. [1]. In that sub-system, the video capturing is performed by using multiple RGB-D sensors [15], which track both the audiovisual and the depth information.

To keep costs and computational load low, the setup considered in this work is based on a capturing system with four RGB-D sensors [29], concretely Kinect v2, which are placed, calibrated and synchronized according to the specifications described in [1]. The four Kinect sensors are connected to four capturing stations, with no particular requirement beyond being able to receive the data from the sensors (e.g. mini PCs). These stations are connected via a Local Area Network (LAN) to a Reconstruction Station with a graphical board supporting GPU (Graphics Processing Unit) operations. In this work, a PC with an Intel Core i7 processor, 32 GB of RAM and a GeForce 1080 Ti board, has been used.

The effective capturing area is approximately a circle with a 3m radius. The RGB-D sensors are placed around the circle and are all pointing towards the action area in the center of the circle. The reconstruction is performed by merging the captured RGB-D frames from each sensor, and extracting their 3D geometry. Then, the data from all sensors are synchronized to achieve a coherent volumetric capturing. After this, a background removal process is performed to isolate the geometry from the color information that is needed for the user's 3D representation. The sensors color information is mapped into voxels, and filtered to remove noise. The *SocialVR* platform supports the representation of the (voxelized) volumetric frames as TVMs.

**4.1.2. Encoding & Transmission**. The reconstructed volumetric videos need be encoded and encapsulated for an appropriate real-time distribution via IP networks. Our *SocialVR* platform supports meshes (e.g. TVMs) for which many compression methods have been proposed (e.g. [27, 43]), and open source compression software solutions are available. In particular, the presented version of the platform has adopted Draco software for the compression of TVMs, and makes use of the open-source RabbitMQ streaming tool [44] for the delivery of the compressed TVMs data.

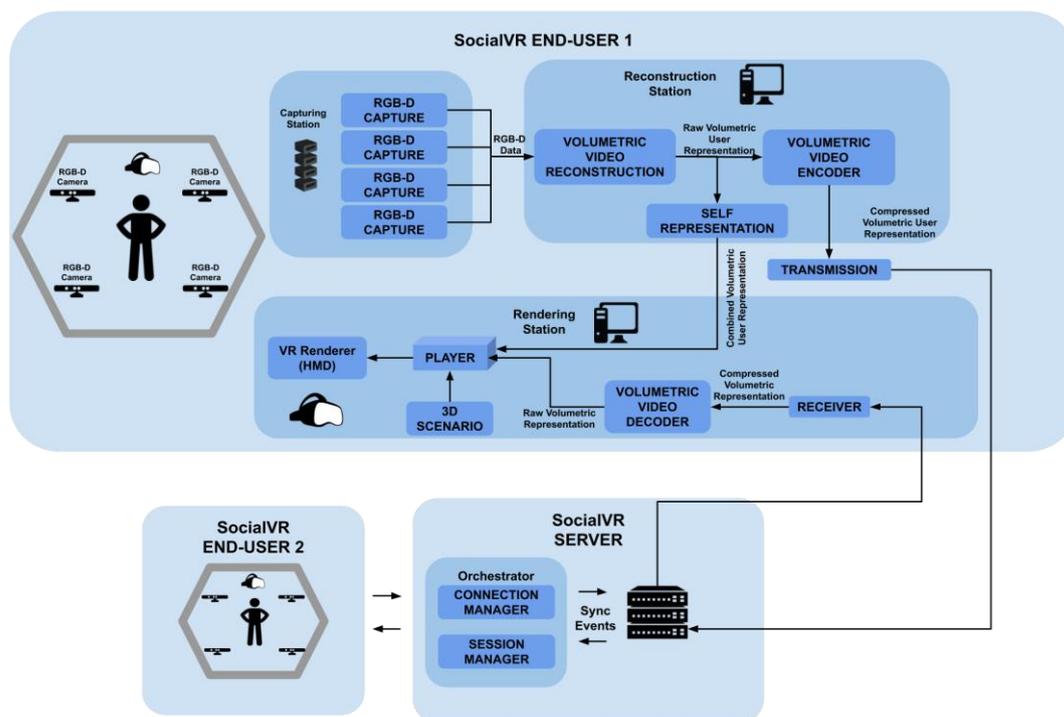

*Figure 6. High-level architecture and flow diagram of the presented SocialVR platform*

**4.1.3. Orchestration**. Orchestration components are commonly used in video conferencing systems to handle the set of audiovisual and control streams [54]. In the presented *SocialVR* platform, an Orchestrator has been developed and integrated to deal with session and stream management tasks. The Orchestrator handles the remote networking information (e.g. IP addresses, ports, protocols…), accommodates all remote users in a shared virtual space, manages the real-time interaction channels, and ensures a consistent synchronized experience.

**4.1.4. Playout**. A Unity-based player has been developed to properly receive, integrate and present all available streams for the end-users' representations and the shared VR scenes. The player includes different engines in charge of:

- Connecting to the Orchestrator to join a shared Social VR session, and exchanging the necessary information to enable interactive and consistent experiences.
- Loading or receiving the 3D virtual scenario where the end-users will be teleported.
- Receiving the data streams for the self and others' representations, as TVMs.
- Seamlessly blending all content formats and streams that constitute the Social VR experience. For traditional streams, the generic audio and video codecs are supported (e.g. [58] [52]).
- Ensuring intra-media, inter-media and inter-device synchronization, in coordination with the Orchestrator.

The player can run on the Reconstruction Station, or on a different station with similar characteristics (see Figure 6). The same station has been used in the setup of this work.

### 4.2. *SocialVR Content Production*

In order to evaluate the potential of Social VR to provide satisfactory shared video watching experiences, an innovative VR story has been produced. The difficulty was to balance an immersive story, which normally limits synchronous interaction [18], with one that facilitates communication between users. The decision, as explained below, was to create a rich narrative experience that includes extra elements to ensure users get insights and converse. This sub-section reviews the content production process of the Social VR content.

**Pre-Production**. After an initial analysis, it was decided to ideate a thriller-like plot revolving around a crime investigation story as the theme for the *SocialVR* experiment. This was expected to provide both commercial relevance and validity for scientific experimentation. Being inspired by movies like *The Usual Suspects*, the ideated VR episode departs from the murder of a celebrity, and revolves around the interrogation to two suspects. The two participants observe the interrogations, playing the role of inspectors. Beyond the VR theme, a number of iterative and interactive design sessions were conducted to assess the most appropriate scenario for telling the story and to recreate the shared environment in which the users will "meet". As the focus of the experiment was the interaction between the two users, and not that much the interaction with the environment and other characters, the decision was to go recreate interrogation scenes behind one-way mirrors, like in classical police stations (see Figures 1.c and 7.a-c).

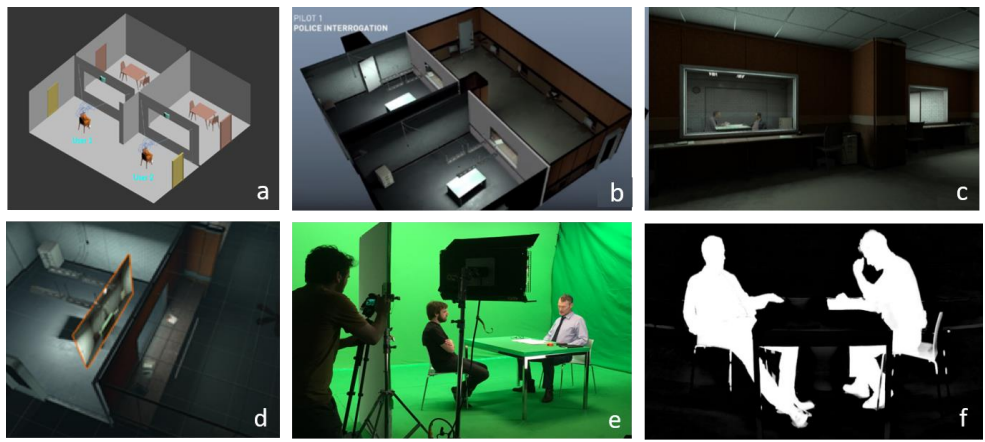

*Figure 7. Created Social VR scenario: (a) Overall view of the modelled 3D environment; (b) Overall view of the recreated 3D environment; (c) Users' viewpoint from the 3D shared interrogation room through different one-way mirrors; (d) 3D scenario with the inserted 2D video billboard for the interrogation scenes; (e) Video shooting over a Chroma key room (f) Example of the masking process*

Likewise, unlike traditional watching apart together scenarios, in which the users watch exactly the same content (as in the protocol test study), it was decided to place the users in a shared observation room, but in front of a different one-way mirror connecting to two separate interrogation rooms (see Figures 1.c and 7.a-c). In each of the separate rooms, a different suspect of the same murder is being interrogated by a police inspector. Therefore, although the users share a common space and can directly see and talk to each other, they can only see and hear one of the two interrogation scenes belonging to the same story. The goal was to boost interaction and the exchange of impressions and findings between the two users to reach a conclusion about the authority of the crime. To stimulate interaction, the following triggers were added: 1) cross-references between the interrogations (the suspects know each other, and that they are being interrogated at the same time); 2) videos from security cameras revealing hints, like inconsistencies in the testimonies; 3) special effects, like light-off situations in the interrogation rooms.

**Script and Casting**. After the selection of the theme and scenario, the next steps consisted of writing the script and casting the actors. The story was further developed, revolving around the murder of Ms. Yelena Armova, a wealthy British celebrity at the peak of her career, in still unknown circumstances. Two persons are the main suspects: Mr. Ryan Zeller, the lover of the victim; and Ms. Christine Gérard, her assistant. The two suspects have a different version about what happened, and the story reflects that they both have things to hide. The two suspects are being interrogated by a police inspector, Sarge. In the *SocialVR* scenario, one user will see and hear the testimony of Ryan Zeller, and the other one of Christine Gérard. Therefore, one script for each interrogation scene was written.

Likewise, a casting process was conducted to select the actors representing the roles of the police inspector and the two suspects. Their participation was necessary, mainly because the objective was to recreate a hyper-realistic VR story including real characters, and not just rely on synthetic avatar-based VR content.

More details about the developed story, the pre-production analysis and tasks, and the casting processes are provided in [14].

**Production and Post-Production**. With respect to content production and consumption, the media formats to use can have direct implications on the required infrastructure, complexity, costs and on the user experience [14]. In order to gain further insights into this, the VR story was produced in three formats: a) *Full 3D version*: 3D environment and 3D-riggered characters animated with Motion Capture (MoCap) techniques; b) *Full 360º version*: all VR content is represented as a (rendered) stereoscopic 360º video; and c) *Hybrid 3D + Billboard version*: a 3D environment with inserted 2D video billboards for dynamic elements, which in this case were the interrogation scenes (see Figure 7.d).

The conducted tests in [14] shown that for this particular Social VR scenario, where the users are sitting and the main actions happen in front of the users and behind a window, the *Hybrid 3D + Billboard version* provides the best experience compared to the other formats, in terms of presence, simulation sickness and QoE. Interestingly, it was found that the *Hybrid 3D + Billboard version* not only provides satisfactory degrees of realism and presence, but also certain levels of motion parallax if the video planes are placed slightly behind the 3D mirror with a slightly bigger size (see Figure 7.d). Based on these results, that *Hybrid 3D + Billboard version* has been adopted in the experiment presented in this paper. Key aspects about its production and post-production process are provided next, but readers can refer to [14] for a detailed description and getting open access links to the created assets.

Regarding the VR environment, the two separate interrogation rooms and the shared space for the two users, together will all associated elements (e.g. chairs, desks), were modelled in 3D, and integrated in a Unity project. The modelled and recreated 3D scenario, resembling a 70's look police station room, can be seen in Figure 7.a-c. The video scenes were shut over a Chrome key room, by using a stereoscopic camera (Canon, with 8-15mm optics sensors, and a separation of 8cm between its lenses). The scene objects, like the table, were also covered in green color. The recording setup can be seen in Figure 7.e.

After the recording and modelling of all assets, post-production processes were conducted for all the raw material, including the required adjustment tasks for an appropriate compositing and seamless blending. Noise reduction and masking processes were conducted for the recorded video billboards (see e.g. Figure 7.f). In addition, color adjustment processes were necessary for an effective removal of the green elements and the replacement with the appropriate color, together with the adjustments to achieve a seamless stereo view. Finally, realistic lighting conditions were recreated in order to provide a natural integration of the users and characters into the 3D virtual environment, and to provide a thriller-like atmosphere. Spatial ambient sound was prepared, coming from the direction of the actions. Subtitles were also produced for each interrogation scene.

**Cost Analysis**. The study in [14] also compares the three content formats in terms of production costs. In particular, without considering the costs associated to the pre-production tasks, participation of actors and software requirements, the production and post-production costs for the *Hybrid 3D + Billboard version* are in the order of 4000 hours over a period of 6 months, by a team of 10 professionals. Approximately, the workload was divided into:

- **CGI** (8 weeks, 1500hours, 4 professionals). This included the 3D modeling of the VR environment and associated elements (chairs, desks, fans...), texturing, illumination and rendering.

- **Video Content** (12 weeks, 1700hours, 4 professionals). This included the Chroma cleaning, color grading, lighting, compositing and integration. The shootings for the two scenes took two days, and require the availability of a professional (stereoscopic) camera and a Chroma key room.

- **Integration in Unity** (5 weeks, 600hours, 2 professionals). This included the integration of all VR content assets and interaction features in Unity, as well as the required adjustments to provide a refined and smooth experience, taking into account the available VR consumption hardware.

- **Direction and Supervision** (24 weeks, 200hours, 1 professional). This included all direction and supervision tasks for the whole content production and post-production processes.

Figure 1.c provides two screenshots of how the whole *SocialVR* scenario looks like, with the integration of the two volumetric captured users, using the TVM technology. A short clip captured from the VE also including the two TVM-based captured users is available at: http://bit.do/socialVR Finally, a video describing the created *SocialVR* content, and summarizing the production process, is available at: https://www.youtube.com/watch?v=aHO5M1qNmjY

## 5. Evaluating the SocialVR Experience

This section presents the two conducted experiments with end-users and professionals to determine the benefits and potential impact of the created *SocialVR* experience.

The results of the protocol test (see Section 3) shown that the experimental protocol, including the Social VR questionnaire, semi-structured interviews and the video recordings, is correctly designed to collect useful data. Such results also suggest that photo-realistic user representations provide a more engaging Social VR experience than cartoon avatars. Participants spent more time talking and looking at each other, and rated the QI and SM higher in the photo-realistic system (VB) than in the avatar system (FS). In the interviews, nearly half of the participants mentioned the importance of facial expressions, natural body movements and hand gestures in assisting communication. More than one third of them mentioned the virtual environment could be more realistic and less blurry.

Based on these insights, we set up an experiment using our *SocialVR* platform (see section 4.1) that enables real-time capturing and rendering of volumetric users representations, including also self-views. By using our platform, users are able see each other's gestures, facial expressions and communicate through audio in real time. They can move their body and hands naturally without the use of controllers. The users experienced the novel produced *SocialVR* content (see Section 4.2).

The following subsections report on the setup and methodology, and on the results of the evaluation with both end-users and professionals.

### 5.1. *Evaluation Apparatus & Methodology*

In the experiment, two users were sitting in two different rooms, but with the same setup and look (see Figure 8). These two rooms were located in different floors of a building, and were interconnected via its networking infrastructure. Each room had the same equipment for the TVM-based end-user's reconstruction, including four RGB-D cameras (Kinects) and five PCs (one per camera plus one controller, Figure 8). TVMs with a resolution of 12k vertices and a frame rate of 7fps were used. As parametrization, we adopted the outcome of the subjective study on mesh compression performance carried by Christaki et al. [8]. A laptop was also used to record the audio and video from each user via its integrated webcam (see Figure 8.b). The rooms had no background noise. Each user was equipped with an Oculus Rift HMD with an integrated microphone and a noise-cancelling headphone for a better perception of the spatial audio. Thus, the users were able to interact through (spatial) audio and (volumetric) visual channels. The users sat in a chair fixed to the floor at the center of the effective capturing region (see Figure 8). An experiment facilitator was present in each room to assist the user and to control the test. The Orchestrator was used to synchronously launch the shared VR experience at both rooms, and chat tools were used to enable communication between the experiment facilitators.

Table 1 lists the main differences between the VB Social VR system [21], used in the protocol test, and the developed *SocialVR* platform, used in the experiment.

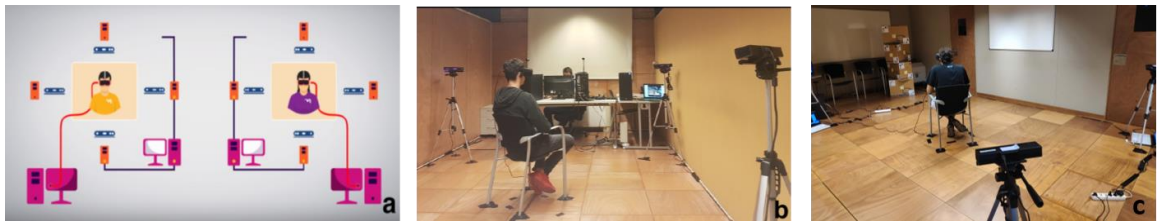

*Figure 8. Experiment Setup: (a) illustration of the apparatus; (b) and (c) photos of the rooms*

Table 1. Comparison between the VB Social VR system and our *SocialVR* platform

| The VB Social VR system | The *SocialVR* platform |
|---:|:---|
| one RGB-D camera | four RGB-D cameras |
| no self-views | self-views |
| 2D representations | 3D representations |
| 360º image for the VR scene | 3D scene |
| watch the same video | watch different, but content-related videos |

### 5.2. *Evaluation with End-Users: Results*

**5.2.1. Procedure**. Fifteen pairs of acquaintance participants were recruited (3f, M=21.6, SD=3.88) for a single test condition using the *SocialVR* platform. After an initial welcome, description of the experiment and filling of consent forms, the users were accommodated. Although the produced *SocialVR* content lasts 8 minutes, the users were allowed to spend more time in the virtual environment after the interrogations for further observation, and in order to exchange their impressions and gathered hints to solve the crime together. After the *SocialVR* experience, participants individually filled in the Social VR questionnaire (see Appendix A, replacing "trailer" by "content" in the question items) and jointly participated in a semi-structured interview. The evaluation was conducted following the test protocol described in Section 3.1.

**5.2.2. Questionnaire Results**. The raw sum of scores (unnormalized scores) of the Social VR questionnaire for the created experience are: PI=39.0(4.0), QoI=42.5(6.8), SM=39.0(5.8). Although technological differences between the VB Social VR system and our *SocialVR* platform exist (see Table 1), and different content pieces were used in their evaluation, the obtained scores in both test conditions are also compared (see Table 2). In the protocol test, the obtained questionnaire results showed that the protocol was appropriate and that the VB Social VR system, which also supports photo-realistic user representations, was shown to perform better than FS and to provide a similar experience as the F2F scenario. Therefore, the goal of such comparison is just to assess whether the subjective scores given to our *SocialVR* platform are comparable with the ones given to the VB Social VR system, which were very satisfactory. Through Wilcoxon-Mann-Whitney test, no significant differences between the two Social VR systems were found in terms of PI, QoI and SM.

**Ad-hoc questions**. The questionnaire included three new Extra Questions (EQ) to evaluate the *SocialVR* content:

- *EQ1: "I liked the created VR contents".*
- *EQ2: "The created VR contents are realistic (i.e. resemble a real scenario)".*
- *EQ3: "The spatiality in the VR scenario (i.e. perceived distances and sizes) is consistent with a real-life scenario".*

These questions had be answered by using a 5-point Likert scale in terms of level of agreement, with the possible answers indicated in Table 3. The distributions of the answers is provided in Table 4, indicating very satisfactory results.

Table 2. Response Scale to the Question Items and their Meaning

| | VB Social VR platform | | | *SocialVR* platform | | |
|---|---|---|---|---|---|---|
| | PI | QoI | SM | PI | QoI | SM |
| Md | 36.0 | 42.0 | 40.0 | 39.0 | 42.5 | 39.0 |
| IQR | 5.0 | 5.0 | 5.5 | 4.0 | 6.8 | 5.8 |

Table 3. Response Scale to the Question Items and their Meaning

| Score | Meaning |
|---|---|
| 1 | Totally Disagree (TD) |
| 2 | Partially Disagree (PD) |
| 3 | Neutral (NN) |
| 4 | Partially Agree (PA) |
| 5 | Totally Agree (TA) |

Table 4. Questions about *SocialVR* content evaluation

| Extra Questions | TD (1) | PD (2) | NN (3) | PA (4) | TA (5) |
|---|---|---|---|---|---|
| EQ1 | 0 | 0 | 2 | 11 | 17 |
| EQ2 | 1 | 0 | 2 | 23 | 4 |
| EQ3 | 0 | 1 | 10 | 11 | 8 |

**5.2.3. Participants' Interaction and Behavior Results**. The participants' interaction and behavior were captured by the laptops' webcams and were also manually annotated by the experimenters. At the beginning of the *SocialVR* experience, the majority of the participants (87%) paid attention to their self-views and greeted each other. While watching the interrogation scenes, participants appeared relaxed and interacted with each other effortlessly. All participants used hand gestures during their conversations. After the interrogations finished, 87% of the participants continued their discussions about the story. Their conversations suggested that they adequately understood the story. Many participants (70%) mentioned that they felt immersed and wanted to stand up and further explore the virtual environment.

**5.2.4. Semi-structured Interview Results**. The open coding method from [53] was applied to analyze the audio transcripts of the semi-structured interviews, as for the protocol test. Since the interviews were conducted with two participants together for each pair, their answers were transcribed and coded as a participant pair, not as individual participants. Therefore, the 15 participant pairs are labelled as P1-P15. The results are summarized into four categories as follows:

**Positiveness towards *SocialVR***. All participants thought the photo-realistic *SocialVR* system enabled them to experience "social presence". They felt "being together" with the other participant, which enriched the overall experience. P9 said, "*I felt being together, there was a connection between us, and this is really an added value to VR. We were exchanging our insights about the interrogation*". P4 mentioned that, "*I felt we were together. I was aware of the activities and feelings of my friend*". All participants also found the VR content (both the environment and the interrogation scenes) immersive and realistic. P11 said, "*The contents were immersive. The distance between me and the objects in the environment was realistic. The furniture seemed real*". P8 mentioned, "*The environment was immersive and realistic. You felt you were part of the story.*

*The policeman and suspects seemed talking to and pointing at you*". The participants also felt comfortable in the virtual environment.

Besides the experience of immersion and social presence, most participants (87%) found the quality of communication satisfactory. They thought that the audio communication and the spatial audio effects had the required quality to enable an immersive experience. The interactions between the two participants were natural. Even though the visual quality for the end-users' reconstruction has room for improvement, being able to see themselves in VR was a fascinating feature for the participants. P12 stated, "*The audio communication was successful and very clear. Visual communication was not so high quality, but it was a fascinating feature to see my full body*". All participants believed such photo-realistic *SocialVR* platform can help maintain important relationships in life. P9 said, "*It is a very innovative and useful solution. We have friends and family members living apart. This would enable us to meet, overcoming distance barriers*".

**Emotion sensing and self-identification in *SocialVR*.** 87% of the participants said they could sense the emotions of their partners in *SocialVR*, mainly based on the audio and gestures of their partners. Some quotes from the participants: "*We were able to feel the emotions, and share the 'WoWs!' (P4)*". "*You can infer the emotions from the audiovisual interactions (P14)*". Eight pairs (53%) pointed out that the VR room was too big. The distance between the two participants was too large for them to see the facial expressions, which makes it difficult to visually perceive the emotions. P7 said, "*We did know the partner was there, but it was a bit far, and facial expressions were blocked by the HMD. So, it was very difficult to tell the emotions from the expressions*".

87% of the participants found the self-identification satisfactory. They also pointed out the limitations of the current self-representation implementation, like noticeable delays when performing gestures and the visual quality. One participant in P3 mentioned, "*I felt self-represented, but the delay for the self-representation sometimes caused confusion when performing gestures*". P7 suggested, "*Identification worked, but it can be much better if the visual quality and delays are improved in the future*". P11, P10, P9 and P4 mentioned that their partner's representation was better than themselves'. One participant in P4 said, "*I felt identified with my representation, but the quality was not high. It was easier to recognize my partner, even though he was wearing the HMD*".

**Missing aspects in *SocialVR*.** Most participants (87%) would like to have an improved visual quality for the users' representations, and more fluid movements (i.e. reduced latency and higher frame rates). 93% of the participants mentioned that they wanted to have multi-sensory experiences, especially haptic feedback. P1, P8 and P12 would like to move freely in VR (e.g. 6 Degrees of Freedom, 6DoF). P12 stated, "*We want to move closer to each other, to see more details of the emotions and gestures*". P4 and P7 suggested adding more interactive experiences rather than only passive watching experiences, like exploring things together. P7 said, "*It was a quite passive experience. We want to perform tasks together, like collaborating with the police or* *my partner, and interacting with the environment*". P3 mentioned that the ability to interact with the VR environment would largely increase the immersion.

**Other use cases and the next generation of *SocialVR***. Participants recommended many other use cases for the photo-realistic *SocialVR* system, including interactive TV/films (73%), collaborative creation (73%), eLearning and eHealth consultation (40%), games (33%), and sports (27%).

The next generation of *SocialVR* is envisioned by participants to blur the boundary between the real and the virtual worlds (P14, P13, P12). P12 imagined "*A 3D space, mixing the real and the virtual worlds, with full details, 6DoF and multi-sensory feedback*". Many participants (47%) foresee a big impact of Social VR in remote social activities, especially at the professional level, such as education, collaborative creation and psychological consultation.

**5.2.5. Reflection on the Results of the *SocialVR* Experiment with End-Users**. From the gathered results of the experiment with end-users, brief reflections can be outlined.

**Added value & Potential**. The results and interview feedback suggest that the *SocialVR* platform provides realistic experiences of two users interacting with each other in real-time, providing satisfactory (self-)identification, perceiving natural body movements and gestures, and enabling to sense emotions. The feedback received from the participants also reflects a positive attitude towards the *SocialVR* platform as a promising new medium to overcome distances, provide presence and togetherness, and enable engaging remote interactions, in a comfortable manner. Furthermore, the participants foresee a big potential in other relevant use cases, like virtual meetings and collaborative scenarios.

**Missing aspects & Suggested Improvements**. The most frequently mentioned missing aspects of the current version of the *SocialVR* platform include: (1) limited visual quality for the user representations, (2) higher fluency and lower latency in the body movements, (3) passive watching experiences with limited interactions, and (4) multi-sensory experiences. These aspects will be the focus of future iterations of the platform and experiences to be produced.

5.3. *Evaluation with Professional Users*

**5.3.1. Procedure**. The Social VR experience was showcased to many recognized VR professionals from a variety of companies and entities, and with different profiles. The goals were mainly: 1) to get feedback about the developed technology and experience from a professional viewpoint; and 2) to explore its potential applicability and business opportunities.

The VR professionals were shown the *SocialVR* experience, by using the same setup explained in Section 5.1. Then, they were asked to fill in a questionnaire and participate in a semi-structured interview. Given that it was not always possible to meet with them at the labs where the *SocialVR* platform was setup, a demo video was produced to still have the chance of getting feedback (https://www.youtube.com/watch?v=Rel5qnj8rxA). If the demo video was shown, few question items from the questionnaire and interviews were obviated, as they were related to the *SocialVR* experience itself.

In total, 25 VR companies participated in the focus groups. 10 of them were based in Spain, 9 in The Netherlands, 3 in Switzerland, 2 in France, and 1 in Greece. The 3 VR companies from Switzerland were the only ones for which the demo video was used. The profiles of the companies/entities and of their representatives was quite diverse: Production of VR experiences (8); VR Exhibitor (1); Research Entities / Academia (2); Service Providers / Broadcasters (3); Government Entities (1); IT Companies (8); and Manufacturers (2). The profiles of their representatives were: CEO/CTO/CCO (9, 2 of them with a non-technical profile); Head/Director of Department (11, 2 of them with a non-technical profile); Product/Business Manager (5, 4 of them with a non-technical profile). All the representatives with a non-technical profile were however very familiar with VR technologies and services.

**5.3.2. Results from the Questionnaire**. This sub-section provides the aggregated results from the questionnaire, clustered by the analyzed aspects. Note that some of the questions were optional, so less than 25 answers could be collected for specific question items. The question items were generally formulated as assertions to be answered by using the 5-point Likert scale in Table 3.

**Potential in the market**. Around 80% of the professionals believe that VR is going to have a big impact in business, with almost 60% having a strong conviction about this. However, there is no general agreement on when that big impact will happen. Almost 20% believe that VR is already having such a big impact, but almost 60% think that it will not happen until three years or even more. Interestingly, all participants agreed (45.5% strongly, 54.5% partially) that the presented *SocialVR* platform can have an impact in the VR sector.

**Benefits of Social VR for "watching videos together"**. The professionals were asked about the most interesting content genres for being watched together with other remote users. The highest ranked genres were: Education, Sports, Culture and Gaming like content.

**Applicability of Social VR**. The professionals were asked about the applicability of Social VR in other use cases. Interestingly, they believe that Social VR is widely applicable, providing even higher benefits than for "watching videos together". As can be seen in Table 5, education, training, virtual meetings and remote collaboration were considered the use cases in Social VR can have the biggest potential. Additional use cases, like Industry 4.0, tourism, and therapy, were also suggested.

**Technological Aspects**. The professionals were asked about the importance of specific technological aspects to build the *SocialVR* platform. These aspects are listed in Table 6, together with the obtained results, which show that most of the included aspects / components were considered important or very important. Note that the last three aspects were not yet implemented, but envisioned for future releases of the *SocialVR* platform. So, they were included in the questionnaire to get feedback about their relevance. The professionals were asked to suggest additional important technological aspects, and the most frequent ones were: scalability (3 professionals), hardware performance (2 professionals), and haptic feedback when interacting with the VR environment (2 professionals).

**Aspects for High Immersion**. The professionals were asked about the importance of a set of aspects to provide high immersion. As can be seen in Table 7, all the aspects included in the question items were considered very important. Likewise, they identified other important aspects, such as providing: multi-sensory stimuli (2 professionals), consistency between the real and virtual worlds (2 professionals), and good illumination (1).

Table 5. Applicability of Social VR

| Use Case | TD(1) | PD(2) | NN(3) | PA(4) | TA (5) |
|---|---|---|---|---|---|
| HealthCare | | 2 | 2 | 4 | 11 |
| Education / Training | | | 1 | 4 | 16 |
| Media / Entertainment | 2 | | 3 | 7 | 10 |
| Meetings | | 1 | 5 | 4 | 12 |
| Exhibitions | | 2 | 4 | 6 | 8 |
| Tele-Work / Remote Collaboration | 1 | 1 | 2 | 7 | 11 |

Table 6. Importance of technological aspects for Social VR

| Technological Aspect | TD (1) | PD (2) | NN (3) | PA (4) | TA (5) |
|---|---|---|---|---|---|
| 3D capture and reconstruction using off-the-shelf cameras | | 1 | 3 | 5 | 2 |
| Distribution of Point Clouds or TVMs | | 1 | 3 | 9 | 7 |
| Spatial audio support | | | 4 | 6 | 11 |
| User self-view | | 3 | 4 | 8 | 6 |
| Live Distribution of Scene Contents | | 2 | 7 | 6 | 6 |
| HMD removal | | 1 | 2 | 4 | 13 |
| Interaction with the VR environment | | 1 | 1 | 10 | 9 |

Table 7. Importance of aspects to provide high immersion

| Aspect | TD (1) | PD (2) | NN (3) | PA (4) | TA (5) |
|---|---|---|---|---|---|
| High Visual quality in the shared VR scene | | | 2 | 5 | 15 |
| High Audio quality in the shared VR scene | | 2 | 4 | 6 | 10 |
| Having a 3D navigable scene | | 1 | 2 | 8 | 10 |
| Having spatial audio | | 2 | 5 | 5 | 10 |
| High Visual quality of the end-users' reconstruction | | 1 | 4 | 6 | 11 |
| Having volumetric (3D) bodies of the end-users | | | 2 | 14 | 4 |

**5.3.3. Results from the Interviews**. The interviews with 25 professionals was audio recorded and transcribed. The open coding analysis from [53] was also applied to sort the transcripts into four main categories, namely: (1) the limitations of the *SocialVR* platform, (2) the quality of interaction and social presence, (3) industrial potentiality, and (4) recommendations. The professionals were labelled as PR1-PR25, and the results for each category are presented next.

**Limitations of the *SocialVR* platform**. Nine professionals pointed out that the current visual quality of the user representation (e.g. frame rate and resolution) and delays make the experience less optimal. Twelve professionals mentioned that the capturing and processing setup is quite complex for commercial adoption. As summarized by PR15, "*Balancing quality and portability would be the biggest challenge on the way to success*". Seven professionals found that hardware in general (e.g. HMDs, controllers…) needs further improvement to ensure comfortable experiences. PR12 mentioned, "*The use has to be non-obtrusive. You have to feel nothing on your head. Maybe, in a few years, people would start accepting to install cameras in their rooms; but right now, the limitations of the hardware make it difficult to calibrate and to reconstruct the scenes*". Finally, two professionals questioned the police interrogation use case to be a rather passive experience. They suggested looking into more interactive *SocialVR* experiences.

**Quality of Interaction and social presence**. Fourteen professionals highlighted that having realistic representations of users, instead of avatars, makes the *SocialVR* platform different compared to other existing commercial Social VR solutions. All professionals stated that they could easily identify each other's representation, and effectively communicated with each other using both verbal and non-verbal cues. They all mentioned that the VR content was professionally created and the virtual environment was realistic. They all agreed that the feeling of social presence and immersion in the virtual space was satisfactory, which is a key aspect for Social VR experiences. PR6 said, "*Having the real representations of users bring a significant added-value for a richer interaction and effective communication*". PR8 also expressed satisfaction about the *SocialVR* platform, "*The achieved quality is comparable to other more expensive solutions, like Microsoft's*".

**Industrial Potential**. All the professionals agreed that Social VR can have a big impact in the industry due to its wide applicability and the potential to support face-to-face-like remote meetings. They all believed that the real-time photo-realistic capturing and reconstruction of users provides an improved quality of interaction, and is a distinctive feature that provides added value. Eleven professionals mentioned that they foresee Social VR having a big impact, especially in the healthcare and training sectors. PR12 forecasted, "*I think healthcare and education in Social VR is going to be massive, and completely change the world*". Still, nine professionals pointed out the risks of adopting the *SocialVR* platform. PR3 stated, "*The SocialVR platform is currently suitable for small-scale showcases, like exhibitions. It is still risky to make business based on this technology for daily tasks in home-based scenarios*".

**Recommendations**. Six professionals recommended to provide free navigable virtual spaces and 6DoF to allow users to explore the virtual space and to interact with virtual artefacts. Five professionals recommended to provide extra sensory experiences, especially with haptic and olfactory stimuli. Five professionals pointed out the importance to see the full faces of users, and suggested to put efforts on the HMDs removal feature. PR6 said, "*Multi-sensory stimuli, like haptic interaction and smells, as well as HMD removal, are missing to resemble a real face-to-face experience*". Four professionals stressed that we should consider the *SocialVR* experience as an integrated user experience, and not merely focus on technological progress. PR12 indicated, "*It is not necessary to focus on one specific aspect, and do it very well. It is better to keep the whole system and experience right. If you want to commercialize the product, you do not need to have everything perfect, but smooth and correct, making sure it delivers benefits to users*". Similarly, PR13 said "*VR is always tricky because there are so many aspects, and all have to work together to provide a satisfactory user experience. The technologies are pretty good now. It is not really just about the sharpness of the content. Integration is the key*".

**5.3.4. Reflection on the Results of the *SocialVR* Experiment with Professionals**. As for the experiment with end-users, brief reflections can be outlined from the experiment with professionals.

**Added value Potential**. The professionals were quite satisfied with the quality of interaction, naturalness and social presence provided by our *SocialVR* platform, with especial emphasis on the real-time realistic representation and integration of users. They generally agreed that the adequate technological aspects were being addressed to provide an innovative and compelling low-cost platform. According to them, the platform has wide applicability and potential to succeed in the market. Beyond shared media consumption, they suggested to deeply look into remote meetings and collaborative scenarios, especially in the training and healthcare sectors.

**Missing aspects / Suggested Improvements**. The limitations identified by the professionals mostly coincide with the ones identified by end-users, but the professionals went a bit further. They raised concerns about the current limitations of available hardware (e.g. cameras, HMDs, controllers) in order to provide high quality, comfortable and non-intrusive experiences, but also to facilitate large-scale deployments in domestic scenarios.

## 6. Discussion

Social VR is a novel communication medium awakening increasing interest of the research community and industry alike. Thus, it becomes crucial to better understand how to qualitatively evaluate and efficiently provide Social VR experiences, together with the potential applicability scenarios and how to successfully achieve market adoption. This paper has shed some light on these previously unexplored research questions.

On the one hand, the paper has proposed a new experimental protocol and applied it in a test based on comparing three shared video watching setups: two state-of-the-art Social VR platforms, and a baseline face-to-face condition. The test not only has served to validate that the protocol worked successfully, but to gather valuable insights about key aspects required to provide satisfactory Social VR experiences (e.g. photo-realistic volumetric representations for the users, high-quality content, effective usage of controllers...). On the other hand, the insights from this initial first test, together with an in-depth analysis of the state-of-the-art, have served to build a new lightweight and low-cost *SocialVR* platform, and produce innovative and appropriate content for this medium. This novel created Social VR experience has been evaluated by both end-users and professionals, using the designed protocol. The obtained results confirm the potential of this medium, gathering very valuable feedback from both the consumers and professional viewpoints.

In particular, the results show highly positive attitude towards Social VR, with the created experience providing satisfactory levels of QoI, SM and PI, as well as a feeling of togetherness while being apart. The results have also served to identify: aspects to be improved and/or added in order to increase the potential of the developed *SocialVR* platform; the applicability scenarios with highest interest and potential; and key factors that will influence the market adoption of Social VR.

Hereafter, key factors, implications and limitations of this study are briefly discussed.

### 6.1. *Limitations and New Requirements*

We are aware of some limitations of the presented exploratory study, like the *SocialVR* experiment not including any baseline condition, and the presented comparison with the VB Social VR system, using different technology and content pieces. However, we believe the results are still very valuable to validate the proposed evaluation protocol, and to confirm the potential of the developed *SocialVR* platform in particular, and of Social VR in general. The participants also identified limitations in terms of technological aspects, like the latency, fluidity and resolution of the end-user's reconstruction (highly influenced by the performance of the associated hardware). However, they generally indicated that the availability of photo-realistic and volumetric representations is a fascinating and distinctive feature that provides added value.

In addition, the results from the interviews helped to identify extra requirements and potential improvements, such as the need for: 6DoF, interaction with the environment, multi-sensory feedback, and HMD removal. These insights are relevant for the community to keep advancing on Social VR, and will also particularly drive the future developments of the presented *SocialVR* platform.

### 6.2. *Evaluation of SocialVR*

An overarching goal of the protocol test and the experiment with end-users has been to propose an evaluation protocol for measuring interaction and immersion in this new medium –*SocialVR*. The protocol includes a set of tools: (1) a Social VR questionnaire, (2) semi-structured interviews to collect subjective user experience data; and (3) a video annotation method to collect user behavior data.

Why do we use a new Social VR questionnaire? Many existing questionnaires have been widely used and validated to measure interaction, presence and immersion experience in both the real world [26, 55] and in VR [17, 22, 47, 59]. However, none of these questionnaires are dedicated for Social VR experiences. That is the reason why we adopted the recently validated Social VR questionnaire from Li et al. [34] for the "watching videos together" scenarios under study. As an example, Figure 4 exhibits results we got by analyzing the questionnaire responses.

Furthermore, by drawing on video annotation techniques, we have shown the possibility of analyzing and comparing users' experiences during the Social VR experiences. This is a first step towards capturing and analyzing Social VR behavior. Future work will focus on exploring other Social VR use cases, collecting further data to validate the Social VR questionnaire. We are aware that objective metrics are also important for evaluation of media services and tools. These include performance metrics (e.g. delays, bandwidth and packet loss, etc.), but also behavior and activity metrics (e.g. gazes, conversational patterns, gestures, movements, etc.). In addition, physiological metrics (e.g. heart rate, Galvanic Skin Response or GSR, etc.) can also contribute to infer the user experience. Although the focus of this study has been on qualitative evaluation via the presented protocol, the above set of evaluation metrics can contribute to profiling the system under study and to achieving a more comprehensive, and even automatic, measurement of the Social VR behavior and experience. This is a relevant research challenge worth to explore in the near future.

### 6.3. *Social Presence*

Social presence belongs to one of the three subcategories of presence (i.e., telepresence (spatial presence), self-presence, and social presence) [32], and it is a crucial aspect for multi-user Social VR systems [41]. Some thoughts and observations about social presence were derived from this study. First, we realized that the context and the user representations are two main factors contributing to the sense of social presence. Spatial presence is part of the contextual factor. As indicated by the users, the sense of "being together" in a realistic and high-quality VR environment, with good audio communication, increased their sense of spatial presence, immersion and social presence. Another contextual factor is about the tasks that the users are engaging with. Some users suggested having more active collaboration tasks, instead of passive watching experiences, would increase the sense of social presence. Physical proximity is a third contextual factor identified by the users. Some of the users in the *SocialVR* experiment mentioned that the large "distance" between them somehow hindered them from seeing their partners' facial expressions. These observations are consistent with the findings in literature (e.g., [23, 34, 41]). Second, the quality of the user representation contributes to the sense of social presence. The quality in this context does not only refer to the visual quality or appearance realism of the representation, but also to the fluency and naturalness of the movements/gestures (behavioral realism). This observation is consistent with the findings in literature (e.g. [17, 41]).

### 6.4. *Technology and Media Formats*

The *SocialVR* platform presented in this paper represents a novel approach for enabling Social VR experiences, by using realistic volumetric representations of users, captured in real-time. The platform includes all the required components, from the capturing of the volumetric videos of the users [1] (represented as TVMs) to their playout in a 3D environment, including the compression, delivery and orchestration processes. An outstanding feature of the developed platform is that its features are provided without requiring highly powerful and expensive equipment, which lowers its cost and eases its deployment. This is a major advantage compared to other existing solutions that require large-scale, complex and expensive infrastructure (e.g. [4, 42]). In terms of technological aspects, the platform will be improved and extended in order to increase its scalability, performance in terms of latency, fluidity and resolution, and provide support for video-based live sources (e.g. broadcasted content).

In addition, the *SocialVR* experiment has included an innovative VR story, connecting two related VR scenes and by seamlessly blending different media formats: a shared 3D environment, stereoscopic video billboards for the interrogation scenes, and 2D videos for the security cameras. That particular combination of media content formats was proved to provide the best user experience, when compared to other combinations of content formats in a previous experiment [14]. Details and insights from the production workflow have been provided to give a general picture of how to provide these novel interactive and immersive, and potentially collaborative, media experiences.

With respect to the users' representations, TVMs have been used. As proved in the literature (Section 2) and in the protocol test (Section 3), realistic representations enhance the user experience compared to the use of avatars. The combination of TVM-based users' representation and the professionally produced VR content in the created *SocialVR* experience has worked successfully. Finally, as mentioned in Section 2, Point Clouds are gaining momentum for the volumetric representation of natural content, and thus the support for Point Clouds in the *SocialVR* platform is planned [25]. Therefore, a comprehensive comparison of content formats for volumetric representation of end-users is also a research study worth to explore.

### 6.5. *Potential and Business Opportunities*

The experiment with end-users has provided initial evidence of the high interest that Social VR awakes. Likewise, the experiment with professional users has not only provided valuable feedback about technological aspects, the created content and the experience itself, but it has also supported the high potential of this medium, providing guidance on where to put the focus, and on the approach to follow to reach a close-to-market, or even market-ready, solution. Valuable feedback about potential business opportunities and use cases with higher interest has been also collected. Finally, the feedback from the professionals encourage continuing the research on this topic, due to its potential impact and the progress achieved so far.

Finally, it should be remarked that the experiments with both end-users and professionals occurred a few months before the COVID-19 times. We are aware the obtained results would have changed if the experiments had been conducted after having suffered the social distancing and travel restriction effects of the pandemic. It is beyond doubt that the need and demand for effective tools for remote communication, interaction and collaboration have tremendously increased in the last few months. Relevant examples are recent works reflecting on the benefits of Social VR for remote conferencing (e.g. [31]). Based on these facts, there is no reason to think about a reduction of the positive feedback and interest gathered with regard to Social VR and our developed platform in current times, but the other way around. This closely relates to the commercial potential and business opportunities.

## 7. Conclusions and Future Work

This paper has presented a novel lightweight *SocialVR* platform that enables users to "meet" and "interact" with each other in a shared virtual environment. The platform allows users who are physically distant to share VR experiences (e.g. video watching, collaboration) in real-time and to communicate through audio and photo-realistic representations of themselves. An experimental protocol was designed and tested, including a Social VR questionnaire, semi-structured interviews, and manual annotations of video recordings. Based on the insights from the protocol test, a *SocialVR* platform enabling photo-realistic user representations was developed, and a professional and innovative VR content was produced. The Social VR experience, combining the platform and content, was evaluated with end-users and professional users.

The evaluation with end-users further validated the experiment protocol, and showed that users had positive ratings about the Social VR experience, mainly in terms of "quality of interaction", "social meaning" and "presence/immersion". They were able to sense each other's emotions and communicate naturally, which well resembled the face-to-face experience. The evaluation with professional users served to confirm the potential impact, as well as to get guidance on: next steps; potential applicability scenarios; and business opportunities. Both experiments confirm the potential of *SocialVR* as a new communication, interaction and collaboration medium.

Due to the novel aspects and the high benefits provided by our *SocialVR* platform, we believe it can be considered as a reference concept and implementation for lightweight, interactive and immersive Social VR experiences, based on photo-realistic volumetric users' representations.

Future work will be targeted at: (1) overcoming mentioned limitations; (2) addressing newly gathered requirements; (3) assessing the impact of the number of users in *SocialVR*; and (4) exploring other *SocialVR* use cases. The research opportunities highlighted in the Discussion section (Section 6) are also worth to explore.

## ACKNOWLEDGMENT

This work has been funded by European Union's Horizon 2020 program, under agreement nº 762111 (VRTogether). Work by Mario Montagud has been additionally funded by the Spanish Ministry of Science, Innovation and Universities with a Juan de la Cierva – Incorporación grant, with reference IJCI-2017-34611.

# Appendix A: Social VR Questionnaire

Please answer the questions, according to your experience of watching the movie trailers.

The scale of the following questions are from 1 to 5, representing the following meanings:

*1 Strongly disagree      2 Disagree      3 Neutral      4 Agree      5 Strongly agree*

| Strongly disagree   1   2   3   4   5   Strongly agree | 1 | 2 | 3 | 4 | 5 |
|---|---|---|---|---|---|
| 1. *"I was able to feel my partner's emotion while watching the trailer."* | ☐ | ☐ | ☐ | ☐ | ☐ |
| 2. *"I was sure that my partner often felt my emotion."* | ☐ | ☐ | ☐ | ☐ | ☐ |
| 3. *"The experience of watching the movie trailer with my partner seemed natural."* | ☐ | ☐ | ☐ | ☐ | ☐ |
| 4. *"The actions used to interact with my partner were similar to the ones in the real world."* | ☐ | ☐ | ☐ | ☐ | ☐ |
| 5. *"It was easy for me to contribute to the conversation with my partner."* | ☐ | ☐ | ☐ | ☐ | ☐ |
| 6. *"The conversation with my partner seemed highly interactive."* | ☐ | ☐ | ☐ | ☐ | ☐ |
| 7. *"I could readily tell when my partner was listening to me."* | ☐ | ☐ | ☐ | ☐ | ☐ |
| 8. *"I found it difficult to keep track of the conversation."* | ☐ | ☐ | ☐ | ☐ | ☐ |
| 9. *"I felt completely absorbed in the conversation."* | ☐ | ☐ | ☐ | ☐ | ☐ |
| 10. *"I could fully understand what my partner was talking about."* | ☐ | ☐ | ☐ | ☐ | ☐ |
| 11. *"I was very sure that my partner understood what I was talking about."* | ☐ | ☐ | ☐ | ☐ | ☐ |

| Strongly disagree   1   2   3   4   5   Strongly agree | 1 | 2 | 3 | 4 | 5 |
|---|---|---|---|---|---|
| 12. *"I often felt as if I was all alone while watching the movie trailer."* | ☐ | ☐ | ☐ | ☐ | ☐ |
| 13. *"I think my partner often felt alone while watching the movie trailer."* | ☐ | ☐ | ☐ | ☐ | ☐ |
| 14. *"I often felt that my partner and I were sitting together in the same space."* | ☐ | ☐ | ☐ | ☐ | ☐ |
| 15. *"I paid close attention to my partner."* | ☐ | ☐ | ☐ | ☐ | ☐ |
| 16. *"My partner was easily distracted when other things were going on around us."* | ☐ | ☐ | ☐ | ☐ | ☐ |
| 17. *"I felt that watching the movie trailer together enhanced our closeness."* | ☐ | ☐ | ☐ | ☐ | ☐ |
| 18. *"Watching the movie trailer together created a good shared memory between me and my partner."* | ☐ | ☐ | ☐ | ☐ | ☐ |
| 19. *"I derived little satisfaction from the trailer watching experience with my partner."* | ☐ | ☐ | ☐ | ☐ | ☐ |
| 20. *"The trailer watching experience with my partner felt superficial."* | ☐ | ☐ | ☐ | ☐ | ☐ |
| 21. *"I really enjoyed the time spent with my partner."* | ☐ | ☐ | ☐ | ☐ | ☐ |

| See graphs below to indicate the emotional closeness | 1 | 2 | 3 | 4 | 5 |
|---|---|---|---|---|---|
| 22. How emotionally close to your partner do you feel now? | ☐ | ☐ | ☐ | ☐ | ☐ |

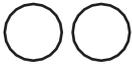 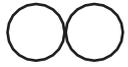 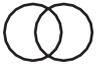 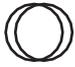

**1**     **2**     **3**     **4**     **5**

| Strongly disagree   1   2   3   4   5   Strongly agree | 1 | 2 | 3 | 4 | 5 |
|---|---|---|---|---|---|

*[Questions 23-28 are not applicable for the real-world condition]*

| | 1 | 2 | 3 | 4 | 5 |
|---|---|---|---|---|---|
| 23. "In the virtual world I had a sense of 'being there'." | ☐ | ☐ | ☐ | ☐ | ☐ |
| 24. "Somehow I felt that the virtual world was surrounding me and my partner." | ☐ | ☐ | ☐ | ☐ | ☐ |
| 25. "I had a sense of acting in the virtual space, rather than operating something from outside." | ☐ | ☐ | ☐ | ☐ | ☐ |
| 26. "My trailer watching experience in the virtual environment seemed consistent with my real world experience." | ☐ | ☐ | ☐ | ☐ | ☐ |
| 27. "I did not notice what was happening around me in the real world." | ☐ | ☐ | ☐ | ☐ | ☐ |
| 28. "I felt detached from the outside world while watching the trailer." | ☐ | ☐ | ☐ | ☐ | ☐ |
| 29. "At the time, watching the movie trailer with my partner was my only concern." | ☐ | ☐ | ☐ | ☐ | ☐ |
| 30. "Everyday thoughts and concerns were still very much on my mind." | ☐ | ☐ | ☐ | ☐ | ☐ |
| 31. "It felt like the trailer watching experience took shorter time than it really was." | ☐ | ☐ | ☐ | ☐ | ☐ |
| 32. "When watching the trailer with my partner, time appeared to go by very slowly." | ☐ | ☐ | ☐ | ☐ | ☐ |

| Extra questions | Yes | No |
|---|---|---|
| 33. Have you watched this movie before? | ☐ | ☐ |
| 34. Did you like the movie trailer? | ☐ | ☐ |

Thank you for your feedback!